\documentclass[%
 reprint,
superscriptaddress,
 amsmath,amssymb,
 aps,
]{revtex4-2}

\usepackage{graphicx}
\usepackage{dcolumn}
\usepackage{bm}

\usepackage[dvipsnames]{xcolor}
\usepackage{braket}

\begin{document}


\title{Magnetically Tuned Continuous Transition from Weak to Strong Coupling in Terahertz Magnon Polaritons}

\author{Andrey~Baydin}
    \email{baydin@rice.edu}
    \affiliation{Department of Electrical and Computer Engineering, Rice University, Houston, Texas 77005, USA}
    \affiliation{The Smalley-Curl Institute, Rice University, Houston, Texas 77005, USA}
\author{Kenji~Hayashida}
    \affiliation{Department of Electrical and Computer Engineering, Rice University, Houston, Texas 77005, USA}
    \affiliation{Division of Applied Physics, Graduate School of Engineering, Hokkaido University, Sapporo, Hokkaido 060-8628, Japan}
\author{Takuma~Makihara}
    \affiliation{Department of Physics and Astronomy, Rice University, Houston, Texas 77005, USA}
\author{Fuyang~Tay}
    \affiliation{Department of Electrical and Computer Engineering, Rice University, Houston, Texas 77005, USA}
    \affiliation{Applied Physics Graduate Program, Smalley-Curl Institute, Rice University, Houston, Texas 77005, USA}
\author{Xiaoxuan~Ma}
    \affiliation{Department of Physics, International Center of Quantum and Molecular Structures and Materials Genome Institute, Shanghai University 200444, Shanghai, China}
\author{Wei~Ren}
    \affiliation{Department of Physics, International Center of Quantum and Molecular Structures and Materials Genome Institute, Shanghai University 200444, Shanghai, China}
\author{Guohong~Ma}
    \affiliation{Department of Physics, International Center of Quantum and Molecular Structures and Materials Genome Institute, Shanghai University 200444, Shanghai, China}
\author{G.~Timothy~Noe~II}
    \affiliation{Department of Electrical and Computer Engineering, Rice University, Houston, Texas 77005, USA}
\author{Ikufumi~Katayama}
    \affiliation{Department of Physics, Graduate School of Engineering Science, Yokohama National University, Yokohama 240-8501, Japan}
\author{Jun~Takeda}
    \affiliation{Department of Physics, Graduate School of Engineering Science, Yokohama National University, Yokohama 240-8501, Japan}
\author{Hiroyuki~Nojiri}
    \affiliation{Institute for Materials Research, Tohoku University, Sendai 980-8577, Japan}
\author{Shixun~Cao}
    \email{sxcao@shu.edu.cn}
    \affiliation{Department of Physics, International Center of Quantum and Molecular Structures and Materials Genome Institute, Shanghai University 200444, Shanghai, China}
\author{Motoaki~Bamba}
    \affiliation{The Hakubi Center for Advanced Research, Kyoto University, Kyoto 606-8502, Japan}
    \affiliation{Department of Physics I, Kyoto University, Kyoto 606-8502, Japan}
    \affiliation{PRESTO, Japan Science and Technology Agency, Kawaguchi 332-0012, Japan}
\author{Junichiro~Kono}
    \email{kono@rice.edu}
    \affiliation{Department of Electrical and Computer Engineering, Rice University, Houston, Texas 77005, USA}
    \affiliation{The Smalley-Curl Institute, Rice University, Houston, Texas 77005, USA}
    \affiliation{Department of Physics and Astronomy, Rice University, Houston, Texas 77005, USA}
    \affiliation{Department of Materials Science and NanoEngineering, Rice University, Houston, Texas 77005, USA}



\begin{abstract}
Depending on the relative rates of coupling and dissipation, a light--matter coupled system is either in the weak- or strong-coupling regime. Here, we present a unique system where the coupling rate continuously increases with an externally applied magnetic field while the dissipation rate remains constant, allowing us to monitor a weak-to-strong coupling transition as a function of magnetic field. We observed a Rabi splitting of a terahertz magnon mode in yttrium orthoferrite above a threshold magnetic field of $\sim$14\,T. Based on a microscopic theoretical model, we show that with increasing magnetic field the magnons transition into magnon polaritons through an exceptional point, which will open up new opportunities for \textit{in situ} control of non-Hermitian systems.
\end{abstract}

\maketitle

Understanding and harnessing the interplay of driving and dissipation in open quantum systems is an important contemporary problem in technology and science. 
Many quantum technologies, including quantum computation, sensing, and transduction, are enabled by coherent light--matter coupling, 
but the coherence can be easily washed out when the matter interacts with dissipative environments. On the other hand, dissipation can be engineered for dissipative quantum error correction schemes as well as for stabilizing qubits against decoherence~\cite{VerstraeteetAl09NP,SchmidtetAl11PRL,ReiteretAl17NC,TouzardetAl18PRX}. Further, driven-dissipative many-body systems can exhibit exotic nonequilibrium phenomena and phases~\cite{HartmannetAl06NP,Foss-FeigetAl17PRL,MaetAl19Nature}.

In a strongly coupled light--matter system, the coupling rate $g$ exceeds the rates of dissipation for light ($\kappa$) and matter ($\gamma$), satisfying $C=4g^2/\kappa\gamma > 1$, where $C$ is called the cooperativity~\cite{PeracaEtAl2020SaS,Forn-DiazetAl19RMP,KockumetAl19NRP}. 
Typically, the matter is placed in a small-mode-volume photonic cavity to enhance $g$.  Such cavity-quantum-electrodynamic systems have recently attracted much theoretical attention as controllable open quantum systems, in which the physics of exceptional points, non-hermitian Hamiltonians, and parity-time symmetry can be explored~\cite{QuijandriaEtAl2018PRA,LuEtAl2021N,MingantiEtAl2019PRA,HuberEtAl2020SP,ArkhipovEtAl2021PRA,PurkayasthaEtAl2020PRR,XieEtAl2021RiP}. In particular, exceptional points, which are spectral singularities where the eigenvalues and eigenvectors coalesce~\cite{Heiss2012JPAMT}, are expected to be useful for manipulating light via nontrivial topological effects~\cite{MiriAlu2019S,OzdemirEtAl2019NM,AshidaEtAl2020AiP}. 

Experimentally, several physical platforms have shown a transition from the weak-coupling regime ($C<1$) to the strong-coupling regime ($C>1$), including intersubband polaritons in quantum wells through gating~\cite{AnapparaetAl05APL} 
or ultrafast optical excitation~\cite{GunteretAl09Nature},
microcavity exciton polaritons in aligned carbon nanotubes through polarization rotation~\cite{GaoetAl18NP}, 
a metal--semiconductor hybrid resonator through spacer thickness variation~\cite{DoironNaik2019AM}, and magnon-polaritons in yttrium iron garnet through position tuning inside a microwave cavity~\cite{ZhangEtAl2017NC,ZhangEtAl2019PRL}. 
However, continuous facile tuning of $g$, $\gamma$, or $\kappa$ by an external field in a single sample has not been achieved.

Here, we demonstrate \textit{in situ} tuning of $g$ by an external magnetic field ($H$) for propagating bulk magnon-polaritons in the antiferromagnetic state of yttrium orthoferrite (YFeO$_3$). We used single-shot terahertz (THz) time-domain spectroscopy~\cite{NoeetAl16OE,BaydinEtAl2021FO} in high magnetic fields up to $30$\,T and observed a field-induced peak splitting of a THz magnon mode in YFeO$_3$. The magnon peak in transmission spectra remains a single peak until a critical field ($\sim$14\,T) is reached, where it splits into two, and the two-peak spectrum persists up to 30\,T. 
Our microscopic model 
quantitatively explains the experimental data. The model shows that the coupling rate $g$ continuously increases with increasing $H$, and exact driving-dissipation compensation ($C = 1$) occurs at the critical field, which is the exceptional point in this system. The magnitude of splitting also increased with increasing sample thickness, in proportion to the square root of the thickness. 

\begin{figure}
    \centering
    \includegraphics{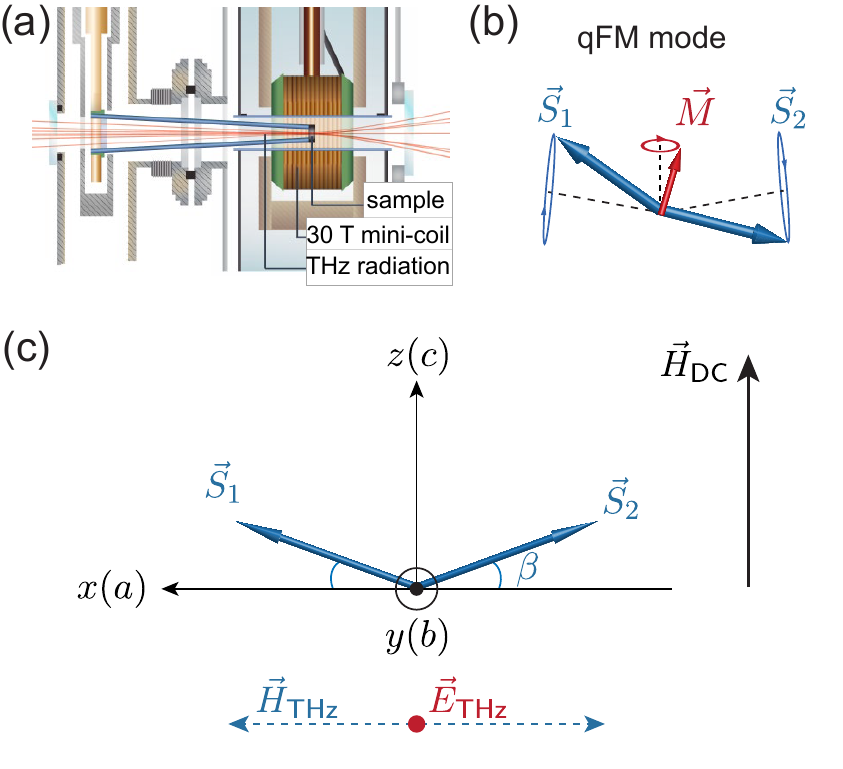}
    \caption{(a)~Schematic of the experimental setup, showing a table-top mini-coil pulse magnet with optical access (RAMBO). (b)~Spin motion of the quasiferromagnetic (qFM) magnon mode in YFeO$_3$. $\vec{S}_1$ and $\vec{S}_2$ are spins in the two respective sublattices with a small canting angle, which produces a net magnetization $\vec{M}$. (c)~Experimental geometry. The two sublattice spins lie in the $a$--$c$ plane with a canting angle $\beta$. The Cartesian coordinates, $x$, $y$, and $z$, are parallel to the crystal $a$, $b$, and $c$ axes, respectively. The electric (magnetic) field component of the THz radiation, $\vec{E}_\mathrm{THz}$ ($\vec{H}_\mathrm{THz}$), was parallel to the $b$-axis ($a$-axis). The black arrow on the right shows the external millimeter-long pulsed magnetic field produced by the RAMBO magnet, which can be considered to be a DC field $\vec{H}_\mathrm{DC}$, during the picosecond-long THz pulse. In this configuration, only the qFM mode is excited through the Zeeman torque.}
    \label{fig:modes}
\end{figure}

The YFeO$_3$ samples we studied were $c$-cut single crystals grown in an optical floating zone furnace. The thickness of the main sample studied was $2.97$\,mm. We performed THz magnetospectroscopy measurements on these samples using the Rice Advanced Magnet with Broadband Optics (RAMBO)~\cite{TayEtAl2022Arxiv}, which combines single-shot THz detection and pulsed high magnetic fields up to 30\,T~\cite{NoeetAl16OE}; see Fig.~\ref{fig:modes}(a). Bursts of THz radiation were generated through optical rectification by passing the output beam of an amplified Ti:Sapphire laser ($1$\,kHz, $150$\,fs, $775$\,nm, $0.8$\,mJ, Clark-MXR, Inc., CPA-2001) through a LiNbO$_3$ crystal. We recorded the time-domain waveform of the THz pulses that transmitted through the sample by electro-optic sampling in ZnTe in a single-shot manner using a reflective echelon~\cite{NoeetAl16OE,BaydinEtAl2021FO}. 

YFeO$_3$ is a canted antiferromagnet and hosts two magnon modes -- the quasiferromagnetic (qFM) mode and the quasiantiferromagnetic (qAFM) mode. The spin motion of the qFM mode is depicted in Fig.~\ref{fig:modes}(b), where $\vec{S}_1$ and $\vec{S}_2$ represent Fe$^{3+}$ spins in the two Fe sublattices, and $\vec{M}$ is the net magnetization as a result of canting. In the present work, we were primarily interested in the coupling between the magnetic field component of the incident THz radiation, $\vec{H}_\mathrm{THz}$, and the qFM mode.
Figure~\ref{fig:modes}(c) schematically shows the experimental configuration we employed. The two spins lie in the $a$--$c$ plane, which are shown by blue solid arrows. The Cartesian coordinates $x$, $y$, and $z$ are parallel to the crystal $a$, $b$, and $c$ axes, respectively. The electric field component of the incident THz radiation, $\vec{E}_\mathrm{THz}$, was along the $b$-axis, and the magnetic field component, $\vec{H}_\mathrm{THz}$, was along the $a$-axis, respectively. In this geometry, only the qFM mode can be excited through the Zeeman torque~\cite{Herrmann63JPCS}. The external millisecond-long pulsed magnetic field generated by the RAMBO system was applied parallel to the $c$-axis; it was essentially constant for the duration of the picosecond-long THz pulse, and therefore, it can be safely considered to be a DC magnetic field, $\vec{H}_\mathrm{DC}$, shown by a black arrow.

\begin{figure}
    \centering
    \includegraphics{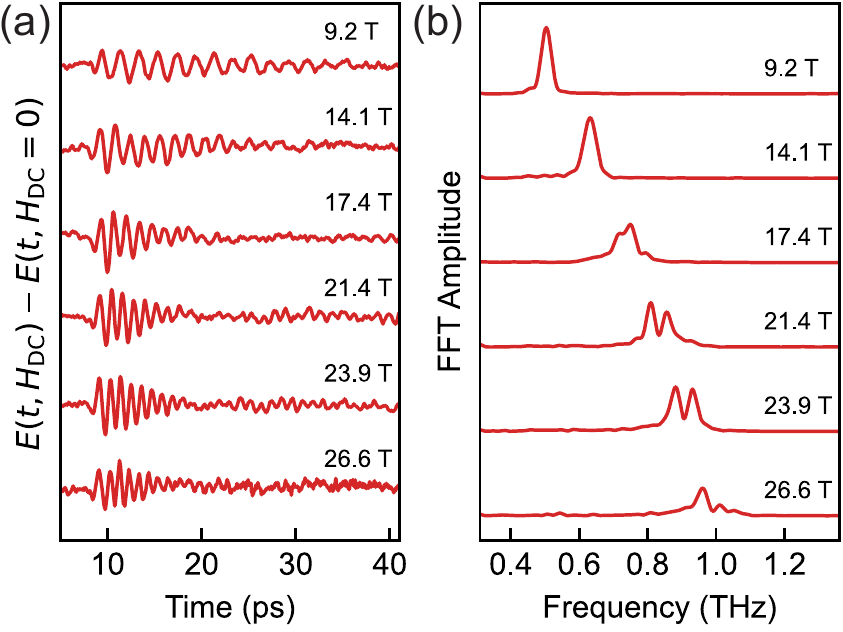}
    \caption{(a)~Magnetic-field-induced changes in the transmitted THz electric field as a function of time at various magnetic fields from $9.2$\,T to $26.6$\,T.  (b)~Fourier transforms of the time-domain waveforms in (a). The traces are vertically offset for clarity both in (a) and (b).}
    \label{fig:waveforms}
\end{figure}

Figure~\ref{fig:waveforms}(a) shows a series of time-domain traces of THz electric fields transmitted through the sample at various selected DC magnetic fields from 9.2\,T to 26.6\,T. For each magnetic field, we subtracted the THz electric-field waveform at zero magnetic field, $E_\mathrm{THz}(t,H_\mathrm{DC}=0)$, from the THz electric-field waveform recorded at the magnetic field, $E_\mathrm{THz}(t,H_\mathrm{DC})$. This differentiation procedure allows us to focus on the magnetic-field-induced changes in the material's THz response. At low magnetic fields ($<14$\,T), we observe long-lived coherent oscillations, with monotonically decaying amplitude, due to the qFM magnon mode excited by the incident THz pulse, as expected for this configuration~\cite{MakiharaEtAl2021NC}. Above 14\,T, however, the data start showing beating behavior, indicating the existence of two oscillation modes with different but similar frequencies. The corresponding Fourier transforms of the data into the frequency domain corroborate this description, as shown in Fig.~\ref{fig:waveforms}(b). At 9.2\,T and 14.1\,T, there is a single peak observed, whose center frequency increases with the magnetic field. Above 14\,T, the peak splits into two, and the splitting magnitude increases with increasing magnetic field. 

To understand our experimental data quantitatively, we developed a microscopic theoretical model. Complete derivations are reported in Supplementary Information. First, we derive the relative permeability in the $c$-cut crystal configuration, following the Herrmann model~\cite{Herrmann63JPCS}. We consider two sublattices with spins $\vec{S}_1$ and $\vec{S}_2$, as depicted in Figs.~\ref{fig:modes}(b) and \ref{fig:modes}(c). The free energy of this system is given by
\begin{equation}
\begin{aligned}
    V = & E\vec{S}_1\cdot\vec{S}_2 - D \left( S_{1x}S_{2z} - S_{2x}S_{1z} \right) - A_{xx} \left( S_{1x}^2 + S_{2x}^2 \right) \\
    & -A_{zz} \left( S_{1z}^2 + S_{2z}^2 \right) - \mu_0\vec{H}_\mathrm{DC} \cdot \left(\vec{S}_1 + {\vec{S}}_2 \right),
    \label{eq:Herrmann_energy}
\end{aligned}
\end{equation}
which contains the isotropic exchange interaction (the first term with coefficient $E$), the Dzyaloshinskii-Moriya interaction (the second term with coefficient $D$), the anisotropy energies (the third and fourth terms with coefficients $A_{xx}$ and $A_{zz}$, respectively), and the Zeeman interaction with the external DC magnetic field (the last term); $\mu_0$ is the vacuum permeability.

The equation of motion for the $i$th component of the spins can be written as
\begin{equation}
    \frac{1}{\gamma}\dot{\vec{R}}_i=\vec{R}_i\times\mathrm{\nabla}_iV-\frac{\alpha}{\gamma_r}{\vec{R}}_i\times{\dot{\vec{R}}}_i,
\label{eq:motion}
\end{equation}
where $\vec{R}_i = \vec{S}_i /  \lvert\vec{S}_i\rvert  $ are the normalized spins with $\lvert\vec{S}_i\rvert=5/2$, 
$\gamma_r$ is the gyromagnetic ratio, and $\alpha$ is the dimensionless Gilbert damping coefficient. By solving Eq.~(\ref{eq:motion}) with the free energy $V$ given by Eq.~(\ref{eq:Herrmann_energy}), we can obtain the resonance frequency, $\omega_0$, and the magnetic susceptibility tensor for the qFM and qAFM modes without any fitting parameters. Because THz magnetic field is along the $a$-axis, we here focus on the $xx$ element of the magnetic susceptibility tensor is shown in Eq.~(\ref{eq:susceptibility}),

\begin{equation}
    \begin{aligned}
        &\chi_{xx} = \frac{\Delta\mu_{xx}}{\omega^{2}_{0} - \omega^{2} -\mathrm{i} \omega 
        \gamma}
        \omega^{2}_{0}, \\ 
        &\Delta\mu_{xx} \propto \frac{\braket{R_{1z} + R_{2z} }^{2}
        }{ \omega^{2}_{0} },
    \end{aligned}
    \label{eq:susceptibility}
\end{equation}
where $\gamma$ is the matter decay rate, which includes Gilbert damping and interactions. 
All necessary physical quantities are taken from the literature (see Supplementary Information). 

\begin{figure}
    \centering
    \includegraphics{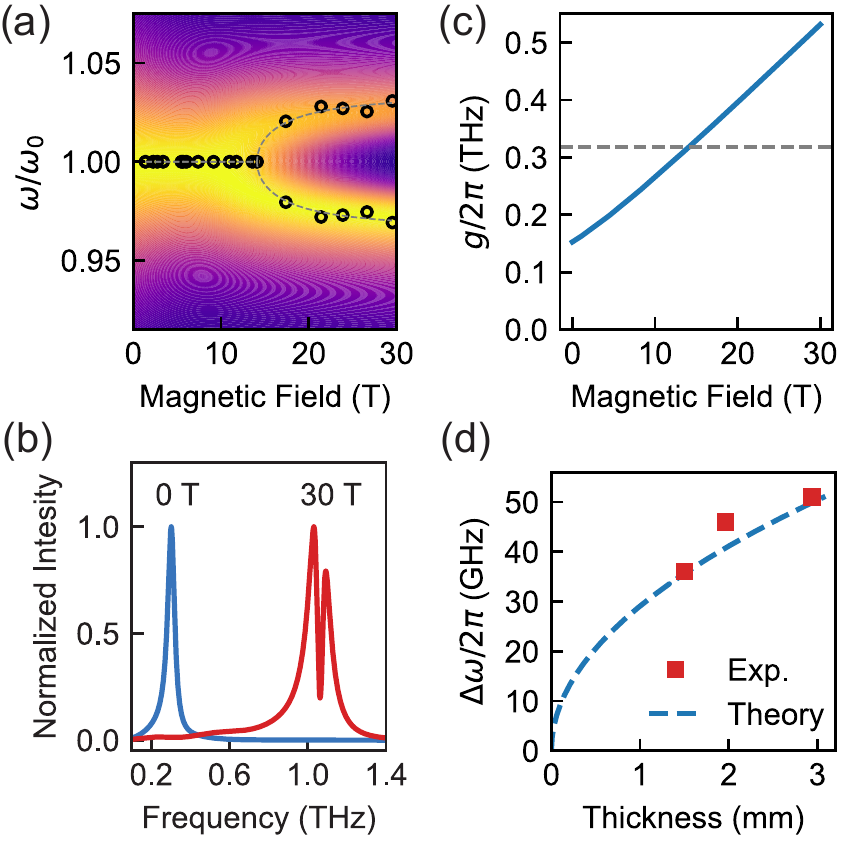}
    \caption{
    (a)~The frequency $\omega$ of the qFM mode, normalized by the center frequency $\omega_0$, as a function of magnetic field. At high fields, the qFM resonance splits into two, which is attributed to the formation of a bulk magnon polariton. The color map is our simulated spectra. The open circles are obtained from experimental data, and the dashed lines are square-root function fits. 
    (b)~Calculated magnon polariton spectra for two values of magnetic field -- $0$\,T and $30$\,T. 
    (c)~Calculated coupling strength $g$ as a function of magnetic field. The dashed line indicates the loss  $\sqrt{\kappa\gamma}/2$. 
    (d)~The magnitude of splitting $\Delta\omega$ versus sample thickness. 
    }
    \label{fig:freq_evolution}
\end{figure}

Next, we calculated THz transmittance spectra by the scattering matrix method using the relative permeability obtained from the microscopic model (see Supplementary Information) and deduced the resonance frequencies. 
Figure~\ref{fig:freq_evolution}(a) summarizes the frequencies ($\omega$) of the peaks normalized by the center magnon frequency ($\omega_0$) determined both from the experiment and calculation. The numerical data is presented as a color map, while open circles indicate the experimental data. Dashed lines are square-root fits.  Figure~\ref{fig:freq_evolution}(b) shows theoretical traces at $0$\,T and $30$\,T obtained from the color map in Fig.~\ref{fig:freq_evolution}(a). Theory and experimental data are in good agreement. As the applied magnetic field increases, the splitting between the two peaks increases. The transition from weak coupling to strong coupling occurs through the exceptional point (a spectral singularity where the eigenvalues and eigenvectors coalesce~\cite{Heiss2012JPAMT}). This dependence of the splitting on the magnetic field can be explained by the magnetic field dependence of the oscillator strength, $\Delta\mu_{xx}$. For the case of scalar susceptibility and zero damping it can be shown (see Supplementary Information) that the frequency splitting, $\Delta\omega$, depends on the oscillator strength, $\Delta\mu_{xx}$ as:
\begin{equation}
    \Delta\omega = \omega_0 \sqrt{\Delta\mu_{xx}}, 
\end{equation}
We see that the oscillator strength and magnon frequency contribute to the frequency splitting of the magnon-polariton. Equation~(\ref{eq:susceptibility}) shows that $\Delta\mu_{xx}$ depends on the average of the $z$ components of the two spins $\langle R_{1z}+R_{2z}\rangle$, which increases continuously with increasing magnetic field applied along the $z$-axis. Thus, the frequency splitting increase with magnetic field. 

With further theoretical analysis, we can obtain the light matter coupling strength, $g$, as a function of the magnetic field, which is plotted in Fig.~\ref{fig:freq_evolution}(c), where the dashed line indicates the light and matter losses, $\sqrt{\kappa\gamma}/2$. The existence of an exceptional point can be naturally explained as the competition between the matter dissipation rate ($\gamma$) and light effective loss rate ($\kappa$), which are independent of the magnetic field, and the light-matter coupling strength ($g$), which increases with the field. At low magnetic fields, the dissipation wins, while at high magnetic fields, the coupling wins. Thus, in this system, the external magnetic field induces an \textit{in situ} continuous transition between the weak-coupling regime (at low fields) and the strong-coupling regime (at high fields). At the exceptional point (14\,T), the loss and the coupling strength are equal. Therefore, the effective loss rate of light can be calculated $\kappa=11$\,THz with $\gamma=0.035$\,THz. Such relatively high effective loss rate (escape rate) of light reflects the absence of the cavity structure.


Finally, we also found that the magnitude of splitting also increases with the sample thickness, $d$; see Fig.~\ref{fig:freq_evolution}(d). A similar splitting of THz magnon polaritons has been previously observed as a function of temperature at zero magnetic field~\cite{GrishuninEtAl2018AP,Shietal2020Arxiv}, 
but only qualitative explanations were given. Here, we demonstrated that the splitting shows a $\sqrt{d}$ dependence at a fixed magnetic field, as shown in Fig.~\ref{fig:freq_evolution}(d). The dashed line, which is proportional to $\sqrt{d}$, was obtained from calculations using our microscopic model (see Supplementary Information), as detailed below, and agrees well with the experimental data. 
The coupling strength, $g$, is known to be proportional to the density of two-level objects, $\rho$, i.e., $g \propto \sqrt{\rho}$~\cite{LietAl18Science,ZhangetAl16NP,PeracaEtAl2020SaS}. Usually, such a dependence is discussed in the context of light-matter coupling in a cavity~\cite{huebl_PRL_2013,Yuanetal2017APL}, where the volume is kept constant, and the density dependence is replaced with the dependence on the number of spins, $g \propto \sqrt{N}$~\cite{Dicke54PR,LietAl18Science,GaoetAl18NP,YahiaouiEtAl2021APP}. In our case, this cannot be true, as the density of spins is constant as a function of thickness. However, the splitting as a function of thickness also follows the square square-root dependence, $\Delta \omega \propto \sqrt{d}$, as summarized in  Fig.~\ref{fig:freq_evolution}(d). Such behavior instead can be understood as arising from the boundary conditions of Maxwell equations for the THz wave propagation, similar to exciton-polaritons in a finite system~\cite{BambaIshihara2009PRB}.

In conclusion, we investigated YFeO$_3$ crystals with different thicknesses using single-shot THz time-domain spectroscopy in high magnetic fields up to 30\,T. We observed that above about 14\,T the quasiferromagnetic magnon mode splits into two peaks and the frequency splitting keeps increasing with increasing magnetic field. This behavior can be explained by the formation of bulk magnon polaritons. Our theoretical model based on the microscopic permeability tensor and scattering matrix method agrees well with the experimental data. From the model, it follows that the frequency splitting dependence on magnetic fields arises from the dependence of the oscillator strength on the magnetic field. Furthermore, we show that the coupling strength can be continuously tuned by the applied magnetic field. Thus, our results demonstrate that the strong photon--magnon coupling can be controlled by the magnetic field. This adds bulk THz magnon polaritons in antiferromagnets to other systems supporting exceptional points, which are promising for further exploration of non-Hermitian physics and advanced sensing. 

\textit{Acknowledgments.} 
J.K.\ acknowledges support from the U.S.\ Army Research Office (grant W911NF2110157). This research was partially supported by the National Science Foundation through the Center for Dynamics and Control of Materials: an NSF MRSEC under Cooperative Agreement No. DMR-1720595. M.B.\ acknowledges support from the JST PRESTO program (grant JPMJPR1767). J.T.\ and I.K.\ acknowledge the support from the Japan Society for the Promotion of Science (JSPS) (KAKENHI No.\ 20H05662). S.C., W.R. and G.M. are grateful for financial support from the National Natural Science Foundation of China (NSFC, No.12074242), and the Science and Technology Commission of Shanghai Municipality (No.21JC1402600). 

\bibliography{main,jun}





\end{document}



\title{Supplementary Information for ``Magnetically Tuned Continuous Transition from Weak to Strong Coupling in Terahertz Magnon Polaritons"}

\author{Andrey~Baydin}
    \email{baydin@rice.edu}
    \affiliation{Department of Electrical and Computer Engineering, Rice University, Houston, Texas 77005, USA}
    \affiliation{The Smalley-Curl Institute, Rice University, Houston, Texas 77005, USA}
\author{Kenji~Hayashida}
    \affiliation{Department of Electrical and Computer Engineering, Rice University, Houston, Texas 77005, USA}
    \affiliation{Division of Applied Physics, Graduate School of Engineering, Hokkaido University, Sapporo, Hokkaido 060-8628, Japan}
\author{Takuma~Makihara}
    \affiliation{Department of Physics and Astronomy, Rice University, Houston, Texas 77005, USA}
\author{Fuyang~Tay}
    \affiliation{Department of Electrical and Computer Engineering, Rice University, Houston, Texas 77005, USA}
    \affiliation{Applied Physics Graduate Program, Smalley-Curl Institute, Rice University, Houston, Texas 77005, USA}
\author{Xiaoxuan~Ma}
    \affiliation{Department of Physics, International Center of Quantum and Molecular Structures and Materials Genome Institute, Shanghai University 200444, Shanghai, China}
\author{Wei~Ren}
    \affiliation{Department of Physics, International Center of Quantum and Molecular Structures and Materials Genome Institute, Shanghai University 200444, Shanghai, China}
\author{Guohong~Ma}
    \affiliation{Department of Physics, International Center of Quantum and Molecular Structures and Materials Genome Institute, Shanghai University 200444, Shanghai, China}
\author{G.~Timothy~Noe~II}
    \affiliation{Department of Electrical and Computer Engineering, Rice University, Houston, Texas 77005, USA}
\author{Ikufumi~Katayama}
    \affiliation{Department of Physics, Graduate School of Engineering Science, Yokohama National University, Yokohama 240-8501, Japan}
\author{Jun~Takeda}
    \affiliation{Department of Physics, Graduate School of Engineering Science, Yokohama National University, Yokohama 240-8501, Japan}
\author{Hiroyuki~Nojiri}
    \affiliation{Institute for Materials Research, Tohoku University, Sendai 980-8577, Japan}
\author{Shixun~Cao}
    \email{sxcao@shu.edu.cn}
    \affiliation{Department of Physics, International Center of Quantum and Molecular Structures and Materials Genome Institute, Shanghai University 200444, Shanghai, China}
\author{Motoaki~Bamba}
    \affiliation{The Hakubi Center for Advanced Research, Kyoto University, Kyoto 606-8502, Japan}
    \affiliation{Department of Physics I, Kyoto University, Kyoto 606-8502, Japan}
    \affiliation{PRESTO, Japan Science and Technology Agency, Kawaguchi 332-0012, Japan}
\author{Junichiro~Kono}
    \email{kono@rice.edu}
    \affiliation{Department of Electrical and Computer Engineering, Rice University, Houston, Texas 77005, USA}
    \affiliation{The Smalley-Curl Institute, Rice University, Houston, Texas 77005, USA}
    \affiliation{Department of Physics and Astronomy, Rice University, Houston, Texas 77005, USA}
    \affiliation{Department of Materials Science and NanoEngineering, Rice University, Houston, Texas 77005, USA}



\begin{abstract}
In this supplemental material, we present a microscopic theory for bulk magnon polaritons using a canted antiferromagnet, YFeO$_3$. We calculate permeability tensor and investigate the electromagnetic wave propagation using scattering transfer matrix.
\end{abstract}

\maketitle

\section{Introduction}
The developed model relies on the microscopic derivation of permeability and predicts magnon frequencies and transmission spectra without the need to invoke any fitting parameters. The light pulse propagation in the bulk magnetic material can be understood as the magnon-polariton, and the coupling between magnon and photon leads to the frequency splitting in the transmission spectra.

The following text is organized as follows. In section~\ref{Sec:Relative_permeability} ``Relative permeability of $\mathrm{YFeO}_{3}$ in the $c$-cut configuration,'' we derive the relative permeability in the $c$-cut configuration of YFeO$_3$ following the Herrmann model~\cite{Herrmann1963JPCS}. We introduce the Gilbert damping term into the equation of motion for a two sublattice model, then we obtain the magnetic susceptibility and relative permeability tensors. Next, in Section~\ref{Sec:transmission} ``Transmission of the qFM mode in the $c$-cut configuration,'' we calculate the transmission spectra of the $\mathrm{YFeO}_{3}$ in the $c$-cut configuration by substituting the relative permeability tensor obtained in the previous section into the Maxwell equations. For the relative permittivity $\varepsilon_{\mathrm{r} }$, we assume that the tensor is diagonal and frequency-independent around the magnetic resonance. 

\bigskip
\section{Relative permeability of $\mathrm{YFeO}_{3}$ in the $c$-cut configuration}
\label{Sec:Relative_permeability}
\subsection{Magnon modes in the two sublattice model}
\label{Subsec : Equation of motion for fluctuation of the two magnetizations}
\begin{figure*}
  \centering
  \includegraphics[width=\textwidth]{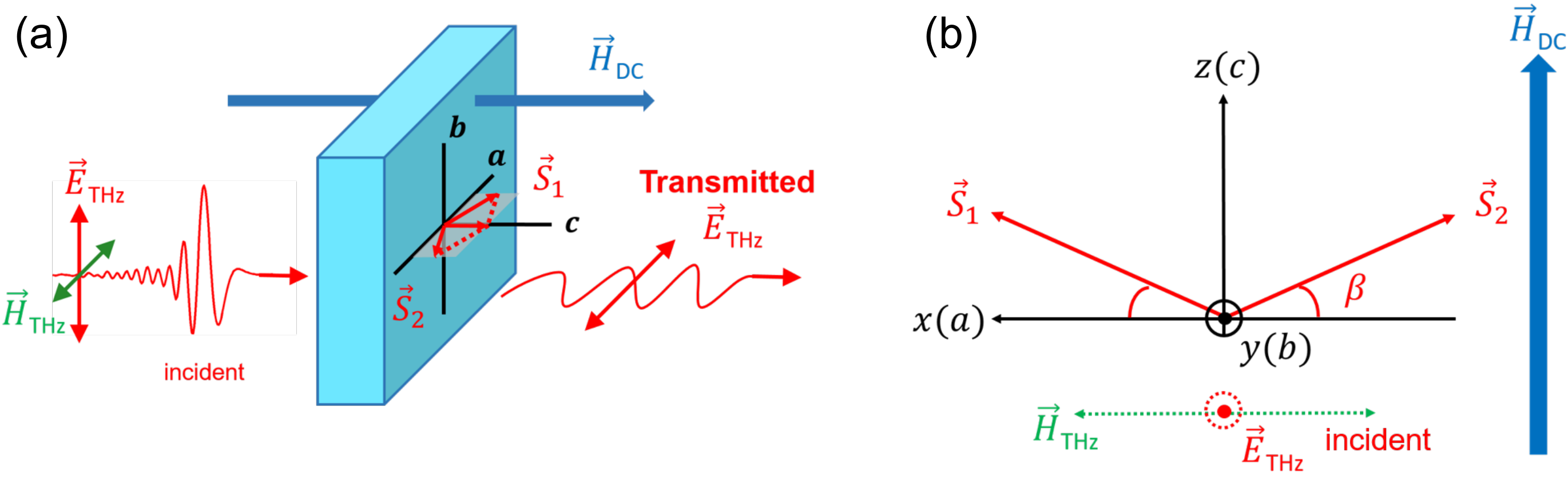}
  \caption{
  (a) The configuration of experimental transmission measurements. Incident THz light is linearly polarized along the $b$-axis. We measured the transmitted light parallel to the $a$-axis. A DC magnetic field is applied to the $c$-axis. (b) Illustration of two spins in the $a-c$ plane. Black solid vectors show Cartesian coordinates, and the red solid arrows show the two spins. $x,y,z$ axes are parallel to the crystal $a,b,c$ axes. Dotted arrows show the incident THz light and thick filled arrow shows the DC magnetic field. $\beta$ is the canting angle. 
  }
  \label{Fig : c-cut_config}
\end{figure*}

In the following calculation, we assume that the system and the magnitudes of the magnetizations are spatially uniform. The magnetic system is characterized by spins of the two sublattices: $\vec{S}_{1}$, $\vec{S}_{2}$. Their magnitude is $S = 5/2$. For simplification, we normalize them as
\begin{align} 
\vec{R}_i & \equiv \vec{S}_i/ \lvert\vec{S}_i\rvert
=
\begin{bmatrix}
X_i & Y_i & Z_i
\end{bmatrix}^{\mathrm{T} }, \lvert\vec{S}_i\rvert = 5/2, \label{Eq : dimensionless magnetization}
\end{align}
where $i = 1, 2$ is the index of the sublattices, T means the transpose operation. The cartesian coordinate system and the configuration of the light propagation in the $c$-cut configuration are depicted in Fig. \ref{Fig : c-cut_config}.

Following Ref.~\cite{Herrmann1963JPCS}, we convert the free energy $F$ to $V=F/g \mu_{\mathrm{B} }S$ in which $g = 2$ is the g-factor, and $\mu_{\mathrm{B} }$ is the Bohr magneton. The free energy, $V$, of the system contains the isotropic exchange interaction $E \vec{R}_{1} \cdot \vec{R}_{2}$, the Dzyaloshinskii-Moriya interaction~\cite{Dzyaloshinsky1958JPCS, Moriya1960PR} $\vec{D} \cdot ( \vec{R}_{1} \times \vec{R}_{2} )$ in which the vector $\vec{D}$ is parallel to the $y$-axis in the present discussion, the single ion anisotropy energy for the $x$-axis $A_{xx} (X^{2}_{1} + X^{2}_{2})$ and the $z$-axis $A_{zz} (Z^{2}_{1} + Z^{2}_{2})$, and Zeeman interactions with an external magnetic field $\vec{H}$. For the exchange interactions, we consider the interactions only between nearest neighbors. For the single-anisotropy, we neglect the $A_{xz}$ term and higher order terms beyond the second order for simplicity of calculation. Note that the exchange interaction is the most dominant among those interactions. Eventually, the free energy of the system is expressed as
\begin{equation}
    \begin{split} 
    V &= E \vec{R}_{1} \cdot \vec{R}_{2} -D (X_{1}Z_{2}-X_{2}Z_{1})\\
    & - A_{xx}(X^{2}_{1}+X^{2}_{2}) - A_{zz}(Z^{2}_{1}+Z^{2}_{2})\\
    & - \mu_{0} \vec{H} \cdot (\vec{R}_{1} + \vec{R}_{2}), 
    \label{Eq : Herrmann energy}
    \end{split} 
\end{equation}
where $\mu_{0}$ is the vacuum permeability. We assume these parameters as
\begin{align} 
E &= \frac{S}{g \mu_{\mathrm{B} } } \times z \times (4.96 \, \mathrm{meV}), \label{EQ : isotropic exchange} \\
D &= \frac{S}{g \mu_{\mathrm{B} } } \times z \times (0.108 \, \mathrm{meV}), \label{EQ : DM interaction} \\
A_{xx} &= \frac{S}{g \mu_{\mathrm{B} } } \times (0.0035 \, \mathrm{meV}), \label{EQ : single ion anisotropy xx} \\
A_{zz} &= \frac{S}{g \mu_{\mathrm{B} } } \times (0.00136 \, \mathrm{meV}), \label{EQ : single ion anisotropy zz} \\
z &= 6, \notag
\end{align}
where $z$ is the number of nearest neighbors. The energy values are taken from literature~\cite{KoshizukaHayashi88JPSJ}.

The magnetic field is set as
\begin{align} 
\vec{H} &= H_{z}\vec{e}_{z} + \vec{h}.  \label{Eq : External magnetic field}
\end{align}
Here, $H_{z}$ is the DC magnetic field and $\vec{e}_{z}$ is the unit vector for $z$-direction and $\vec{h}$ represents the magnetic component of the THz pulse.

The equation of motion for spins is
\begin{align} 
    \frac{1}{\gamma} \dot{ \vec{R} }_i = \vec{R}_i \times \nabla_i V -\frac{\alpha}{\gamma} \vec{R}_i \times \dot{ \vec{R} }_i, \label{Eq : equation of motion}
\end{align}
where $i = 1, 2$, $\gamma$ is the gyromagnetic ratio, and $\alpha$ is the dimensionless Gilbert damping coefficient.

To obtain equilibrium positions of the spins $\braket{\vec{R}_{1}} = \left(\cos \beta, 0, \sin \beta \right)$ and $\braket{\vec{R}_{2}}= \left(-\cos \beta, 0, \sin \beta \right)$ with a canting angle $\beta$ as shown in Fig.~\ref{Fig : c-cut_config}, we solve the equation of motion Eq.~(\ref{Eq : equation of motion}) for $\dot{ \vec{R} }_i = \vec{0}$. Because the isotropic exchange interaction has the largest energy in magnetic interactions, the canting angle is approximated to
\begin{align} 
    \tan 2\beta = \frac{D + \mu_{0} H_{z} }{E+A_{xx}-A_{zz} }.
\end{align}

Spins can be separated into equilibrium positions and fluctuation terms. Applying a linear approximation on the fluctuation $\delta \vec{R}_i = \vec{R}_i - \braket{\vec{R}_i}$ in Eq.~(\ref{Eq : equation of motion}), we can obtain a set of linear differential equations. In the absence of the relaxation term $\alpha$~\cite{Herrmann1963JPCS}, the qFM frequency, $\omega_{\mathrm{FM} }$, is
\begin{align} 
    \left( \frac{\omega_{\mathrm{FM} } }{\gamma} \right)^{2} & = (b - a) (d + c), \label{Eq : Herrmann2subFMmode}
\end{align}
and the qAFM frequency, $\omega_{\mathrm{AFM} }$, is
\begin{align} 
    \left( \frac{\omega_{\mathrm{AFM} } }{\gamma} \right)^{2} & = (b + a) (d - c). \label{Eq : Herrmann2subAFMmode}
\end{align}
The coefficients are
\begin{equation}
\begin{split} 
    a &= -[E \cos 2\beta + D \sin 2\beta\\
    & \quad + 2 (A_{xx} \cos^{2} \beta + A_{zz} \sin^{2} \beta) + \mu_{0} H_{z} \sin \beta ],\\
    b &= E,  \\
    c &= E \cos 2\beta + D \sin 2\beta + 2 (A_{xx}-A_{zz}) \cos 2\beta\\
    &\quad + \mu_{0} H_{z} \sin \beta,
    \label{Eq : Herrmann c} \\
    d &=-(E \cos 2\beta + D \sin 2\beta). 
\end{split} 
\end{equation}

\bigskip
\subsection{Relative permeability for the magnetic resonance}
From the equation of motion in Eq.~(\ref{Eq : equation of motion}), we can obtain the magnetic susceptibility for the qFM and qAFM mode~\cite{Herrmann1963JPCS}. 
From the Fourier transform of the equation, the fluctuation of weak magnetization $\delta \vec{m} = M ( \delta \vec{R}_{1} + \delta \vec{R}_{2}  )$ is expressed as
\begin{empheq}[left=\empheqlbrace]{align} 
\begin{aligned}
    \delta m_{x} & = \frac{2 \gamma^{2} M  }{\omega^{2}_{\mathrm{FM} } - \omega^{2} -\mathrm{i} \omega \gamma (b-a) \alpha} \sin \beta \left[  (b-a)\sin \beta h_{x} + \frac{\mathrm{i}\omega }{\gamma} h_{y}  \right],  \\
    \delta m_{y} & = \frac{2 \gamma^{2} M}{\omega^{2}_{\mathrm{FM} } - \omega^{2} -\mathrm{i} \omega \gamma (b-a) \alpha} \left[ -\frac{\mathrm{i}\omega }{\gamma}\sin \beta h_{x} + (d+c)h_{y} \right],  \\
    \delta m_{z} & = -\frac{2 \gamma^{2} M}{\omega^{2}_{\mathrm{AFM} } - \omega^{2} +\mathrm{i} \omega \gamma (b+a) \alpha} \cos^{2} \beta(b+a)h_{z}.
\end{aligned}
\end{empheq}
$M$ is a unit of magnetization and set as $\mu_{0}\times (5 \mu_{\mathrm{B} }/2)/V_{2\mathrm{sub} }$, where $V_{2\mathrm{sub} }$ is the volume for two sublattices. The value of $V_{2\mathrm{sub} }$ is set as 224/8$\text{\r{A}}^{3}$~\cite{HahnetAl14PRB}.
Since we assume $E \gg D, A_{xx}, A_{zz}$ and the sum of the equilibrium spins is approximated as $R_{1z} + R_{2z} = 2\sin \beta \simeq (D+H_{z})/E$, the fluctuation can be approximated as
\begin{align} 
\delta \vec{m} & = \dvec{\chi } \vec{h}, \\
\begin{bmatrix}
\delta m_{x} \\
\delta m_{y} \\
\delta m_{z} \\ 
\end{bmatrix}
&=
\begin{bmatrix}
 \chi_{xx} & \chi_{xy} & 0 \\
 \chi_{yx} & \chi_{yy} & 0 \\
 0 & 0 & \chi_{zz} \\
\end{bmatrix}
\begin{bmatrix}
  h_{x} \\
  h_{y} \\
  h_{z} \\
\end{bmatrix}, 
\end{align}

Since we analyze the high field splitting of the qFM mode, the $zz$ component relevant to the qAFM mode is not necessary for our purpose. From this susceptibility tensor, we can obtain the relative permeability tensor $\dvec{\mu}_r = \hat{1} + \dvec{\chi }$, where $\hat{1}$ means the 3-by-3 identity matrix.
\begin{equation}
  \left\{
    \begin{aligned}
    & \chi_{xx} = \frac{\Delta \mu_{xx}  }{\omega^{2}_{\mathrm{FM} } - \omega^{2} -\mathrm{i} \omega \gamma (b-a) \alpha} \omega^{2}_{\mathrm{FM} }, \quad \Delta \mu_{xx} = \frac{\gamma M \cdot \gamma E  }{ \omega^{2}_{\mathrm{FM} } } \braket{R_{1z} + R_{2z} }^{2},  \\
    & \chi_{yy} = \frac{\Delta \mu_{yy}  }{\omega^{2}_{\mathrm{FM} } - \omega^{2} -\mathrm{i} \omega \gamma (b-a) \alpha} \omega^{2}_{\mathrm{FM} },  \quad \Delta \mu_{yy} =  \frac{M}{E},  \\
    & \chi_{xy} = - \chi_{yx} =\frac{ \mathrm{i} \Delta \mu_{xy}  }{\omega^{2}_{\mathrm{FM} } - \omega^{2} -\mathrm{i} \omega \gamma (b-a) \alpha }\omega \gamma M, \quad \Delta \mu_{xy} =   R_{1z} + R_{2z}, \label{EQ : chi_xy} \\
    & \chi_{zz} =\frac{\Delta \mu_{zz}  }{\omega^{2}_{\mathrm{AFM} } - \omega^{2} +\mathrm{i} \omega \gamma (b+a) \alpha} \omega^{2}_{\mathrm{AFM} }, \quad \Delta \mu_{zz} =  \frac{M}{E}.
  \end{aligned}
  \right.
\end{equation}

\bigskip
\section{Transmission of the qFM mode in the $c$-cut configuration}
\label{Sec:transmission}
In this section, we calculate the transmission spectra of the $\mathrm{YFeO}_{3}$ in the $c$-cut configuration using the relative permeability obtained in the previous section. For the relative permittivity $\varepsilon_r$, we assume that the tensor is diagonal and frequency-independent around the magnetic resonance. We set the transmission geometry as normal incidence for simplicity, and the polarization of incident light is linear along the $b$-axis (Fig. \ref{Fig : c-cut_config}). We calculated the transmitted light parallel to the $a$-axis. Because the magnetic field of the incident THz pulse is parallel to the $a$-axis and the sample configuration is the $c$-cut (Fig. \ref{Fig : c-cut_config}), a Zeeman torque from the incident light can only excite the qFM mode. Generally, the transmitted light influenced by the qFM resonance will become elliptical. Therefore, we take into account the polarization degree in our calculation.

For the calculation, we use the scattering matrix method. As a sign convention, we use $e^{+i(\vec{k} \cdot \vec{r}- \omega t )}$, the complex relative permittivity $\varepsilon_r = \varepsilon_{1}+ i \varepsilon_{2}, \varepsilon_{2}>0$, and the complex refractive index $N = n +  i \kappa, \kappa>0$. Note that $\mathrm{YFeO}_{3}$ is a birefringent material and the diagonal permittivity tensor has different values for each component.

The calculation starts from Maxwell equations in the medium with the following permittivity and permeability tensors:
\begin{align} 
\dvec{\varepsilon}_r&=
\begin{bmatrix}
\varepsilon_{a} & 0 & 0 \\
0 & \varepsilon_{b} & 0 \\
0 & 0 & \varepsilon_{c} \\
\end{bmatrix},
\dvec{\mu}_r =
\begin{bmatrix}
 \mu_{aa} &  \mu_{ab} & 0 \\
 \mu_{ba} &  \mu_{bb} & 0 \\
0 & 0 &  \mu_{cc} \\
\end{bmatrix}.
\end{align}
$\dvec{\varepsilon}_r$ is a background permittivity, $\dvec{\mu}_r=\hat{1} + \dvec{\chi }$  
and $\dvec{\chi }$ comes from Eq.~(\ref{EQ : chi_xy}). After the Fourier transform and considering the normal incidence, we can obtain wave equations in the frequency domain as
\begin{align} 
\frac{\diff }{\diff z^{\prime} }
\begin{bmatrix}
E_{x} \\
E_{y} \\
\tilde{H}_{x} \\
\tilde{H}_{y} \\
\end{bmatrix}
&=
\begin{bmatrix}
0 & 0 & \mu_{ba} & \mu_{bb} \\
0 & 0 & -\mu_{aa} & -\mu_{ab} \\
0 & \varepsilon_{b} & 0 & 0 \\
-\varepsilon_{a} & 0 & 0 & 0 \\
\end{bmatrix}
\begin{bmatrix}
E_{x} \\
E_{y} \\
\tilde{H}_{x} \\
\tilde{H}_{y} \\
\end{bmatrix}.
\end{align}
$E_i$ and $\tilde{H}_i$($i = x,y$) are, respectively, the electric and reduced magnetic field defined as
\begin{align} 
\vec{ \tilde{H} }=+i \sqrt{\frac{\mu_{0} }{\varepsilon_{0} } } \vec{H},
\end{align}
where $\varepsilon_{0}$ is the permittivity of vacuum. In addition, we normalized $z$ in the form of $z^{\prime}=k_{0}z$ with a vacuum wavenumber $k_{0}=\omega/c$ where $c$ is the speed of light.

The background relative permittivity is given by the refractive indices ($n_j, j=a,b$) and extinction coefficients($\kappa_j, j=a,b$) through the following equation
\begin{align} 
\varepsilon_j = \left(n^{2}_j - \kappa^{2}_j\right)+ i2 n_j\kappa_j, \quad j=a,b.
\end{align}
For simplicity, we assume $n_j, \kappa_j$ are independent of frequency and magnetic field. The refractive indices for the $a$ and $b$ axes are $n_{a} + i \kappa_{a} = 4.55 + 0.024 i $ and $n_{b} + i \kappa_{b} = 4.35 + 0.02 i$, respectively, which are in agreement with the literature~\cite{Jinetal2013PRB}. The values of magnetic interactions are shown in Eq.~(\ref{EQ : isotropic exchange}-\ref{EQ : single ion anisotropy zz}). The Gilbert damping term, $\alpha$, is obtained as $\alpha=\Delta\nu/\gamma H_\mathrm{E}=9.8\times10^{-4}$ with the exchange field, $H_\mathrm{E}=640$\,T~\cite{LinEtAl2015APL} and measure qFM linewidth, $\Delta\nu=35$\,GHz. The obtained value is close to previously reported~\cite{KimEtAl2017SR}

Through Maxwell's boundary conditions, the electromagnetic fields in a single layer can be connected with the adjacent layers. We derived a scattering matrix equation for expressing out-going waves by incoming (incident) waves. As initial conditions, we set the incident wave as the experimental waveform.
\begin{figure}
    \centering
    \includegraphics[width=\textwidth]{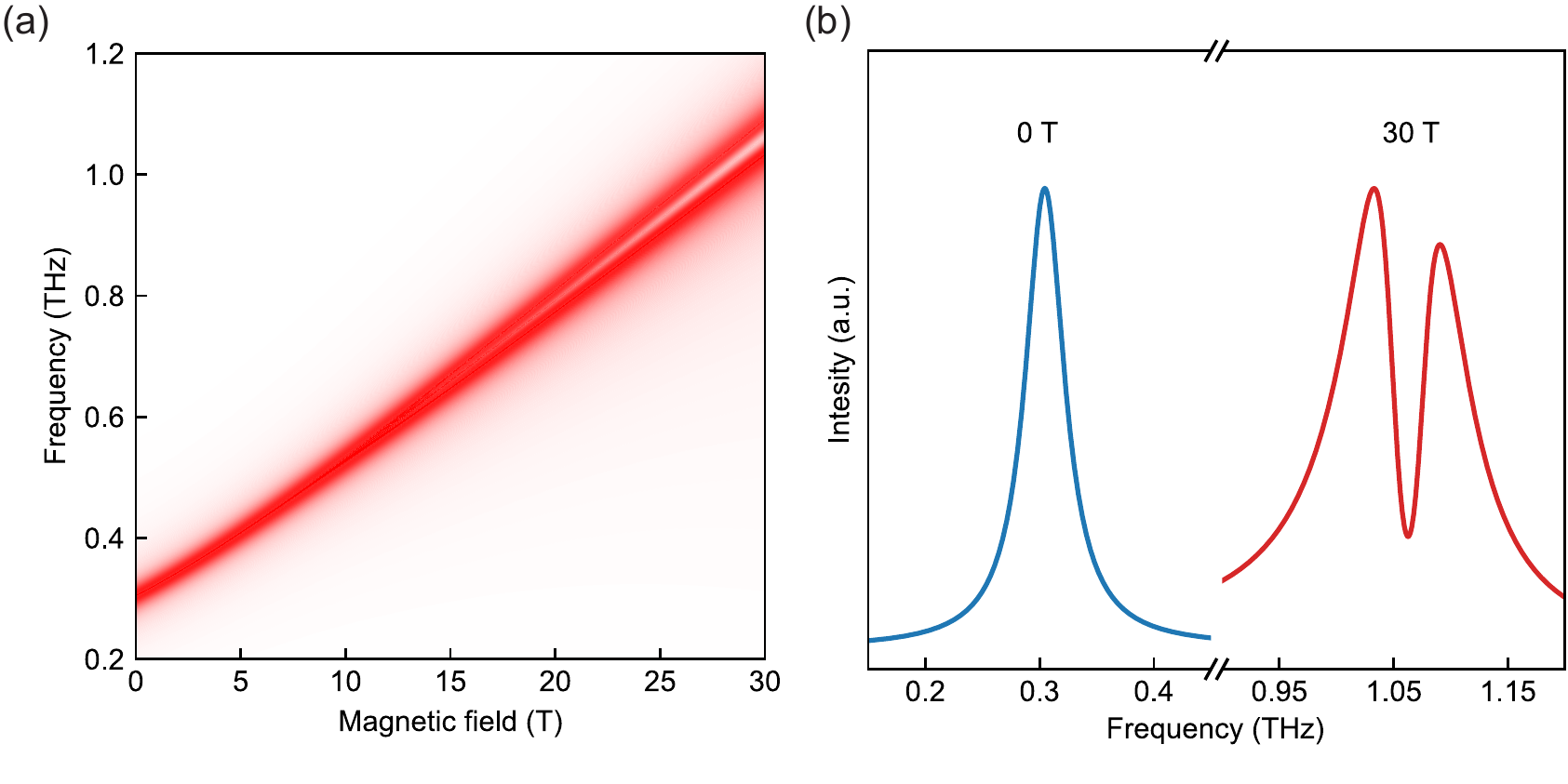}
    \caption{
    (a) Transmittance map, normalized by the maximum value at each magnetic field. The white area shows low transmission and the red area show high transmission. 
    (b) Normalized transmittance at 0\,T and 30\,T from (a). 
    }
    \label{Fig:transmission}
\end{figure}

The calculated transmission color map is shown in Fig.~\ref{Fig:transmission}(a) and its slices at $\mu_{0} H_{z}=0$~T and $\mu_{0} H_{z}=30$~T are shown in Fig.~\ref{Fig:transmission}(b). As magnetic field increases, the single peak corresponding to qFM resonance splits into two peaks. This splitting is an indication of formation of bulk magnon polariton. 

The magnetic-field dependence of the splitting can be explained by that of the oscillator strength. The dispersion relation of the magnon-polariton is determined from Maxwell equations
\begin{empheq}[left=\empheqlbrace]{align} 
\nabla \times \vec{E} &= - \frac{\partial  }{ \partial t} \vec{B}, \, \nabla \times \vec{H}= \varepsilon_{0} \frac{\partial }{ \partial t} \vec{E}, \\
\nabla \cdot \vec{E} &= 0, \quad \nabla \cdot \vec{B} = 0, \\
\vec{B} &= \mu_{0} \dvec{ \mu }_r \vec{H}, \quad \dvec{ \mu }_r = \dvec{1} + \dvec{ \chi }.
\end{empheq}
This gives the following wave equation
\begin{align} 
\nabla^{2} \vec{E} &= \frac{1 }{c^{2} } \dvec{ \mu }_r \frac{\partial }{\partial t} \vec{E}.  \label{EQ : wave equation}
\end{align}
Substituting $\vec{E} = \vec{E}_{0} \exp \left[ i(\vec{k} \cdot \vec{r}-\omega t ) \right]$ into Eq.~(\ref{EQ : wave equation}), the equation is rewritten as
\begin{align} 
\left( \frac{  c^{2} k^{2} }{  \omega^{2} } \dvec{1} - \dvec{ \mu }_r \right) \vec{E} &= \dvec{0}. \label{EQ : FT wave equation}
\end{align}
The determinant of the coefficient matrix in Eq.~(\ref{EQ : FT wave equation}) gives an equation for the dispersion relation,
\begin{align} 
\begin{vmatrix}
\frac{  c^{2} k^{2} }{  \omega^{2} } -1 - \chi_{xx} & - \chi_{xy} \\
- \chi_{yx} & \frac{  c^{2} k^{2} }{  \omega^{2} } -1 - \chi_{yy} \\
\end{vmatrix}&=0. \label{EQ : equation for dispersion relation}
\end{align}

Although the magnon-polariton frequencies can be obtained by solving this equation, for simplicity, we suppose that the susceptibility is expressed as a scalar function with zero damping $\alpha = 0$, i.e., $\chi_{xx}=\chi_{yy}=\chi$ and $\chi_{xy}=\chi_{yx}=0$, as
\begin{align} 
\chi &= \frac{ \Delta \mu \omega^{2}_{\mathrm{FM} } }{ \omega^{2}_{\mathrm{FM} } -\omega^{2} }.  \label{EQ : scalor chi}
\end{align}
Here, we call $\Delta\mu$ as an oscillator strength of the qFM magnon. In this scalar susceptibility case, Eq.~(\ref{EQ : equation for dispersion relation}) is transformed into
\begin{align} 
\omega^{4} - \left( c^{2}k^{2} + \omega^{2}_{\mathrm{FM} } + \Delta \mu \omega^{2}_{\mathrm{FM} } \right) \omega^{2} + c^{2}k^{2} \omega^{2}_{\mathrm{FM} } &=0. 
\end{align}
The solution of this equation gives
\begin{equation}
\begin{split} 
    \omega^{2}_{\pm} &= \frac{1}{2} \biggl( c^{2}k^{2} + \omega^{2}_{\mathrm{FM}} + \Delta \mu \omega^{2}_{\mathrm{FM}} \biggr.\\
    & \biggl. \pm \sqrt{ \left( c^{2}k^{2} + \omega^{2}_{\mathrm{FM} } + \Delta \mu \omega^{2}_{\mathrm{FM} } \right)^{2} - 4c^{2}k^{2} \omega^{2}_{\mathrm{FM}}} \biggr).
\end{split}
\end{equation}
The difference $\omega^{2}_{+}-\omega^{2}_{-}$ is minimized at $c^{2}k^{2}=(1 - \Delta \mu ) \omega^{2}_{\mathrm{FM} }$. At this wavenumber, the two magnon-polariton frequencies are expressed as
\begin{align} 
\omega^{2}_{\pm} &= \omega^{2}_{\mathrm{FM} } \left( 1 \pm \sqrt{\Delta \mu} \right).
\end{align}
Then, the splitting $\Delta \omega = \omega_{+} - \omega_{-}$ is obtained as
\begin{align} 
\Delta \omega &= \omega_{\mathrm{FM} } \sqrt{ \Delta \mu}. \label{EQ : approximated deviation in a final form}
\end{align}
In this way, the oscillator strength and magnon frequency contribute to the splitting in the magnon-polariton spectra. 
\begin{figure} 
  \centering
  \includegraphics[width=0.5\textwidth]{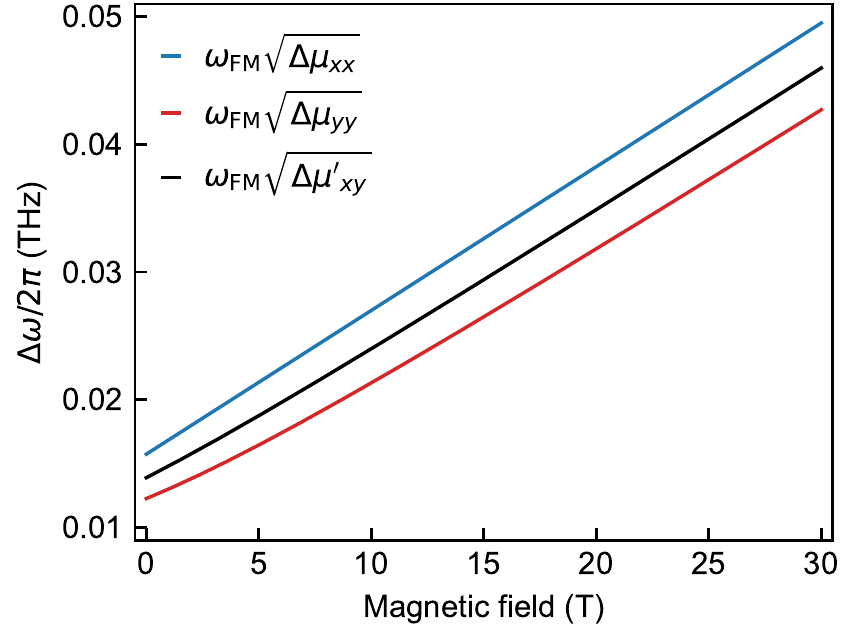}
  \caption{The square root of the denominators of the susceptibility tensor $\omega_{\mathrm{FM} } \sqrt{\Delta \mu_{xx} }$, $\omega_{ \mathrm{FM} }\sqrt{\Delta \mu_{yy} }$, and $\omega_{ \mathrm{FM} } \sqrt{\Delta \mu^{\prime}_{xy} }$ are plotted as functions of the external DC magnetic field. $\Delta \mu^{\prime}_{xy}$ is defined in Eq. (\ref{EQ : muxy^prime}). These values correspond to the frequency splitting in the case of scalar susceptibility.}
  \label{Fig : splitting with muxx muyy muxy}
\end{figure}

In the tensor case as shown in Eq.~(\ref{EQ : equation for dispersion relation}), while it is not easy to derive an analytical expression of the splitting, it is easily understood that the numerators of the susceptibility tensor in Eq.~(\ref{EQ : chi_xy}) appear in the splitting as in the scalar case. In Fig.~\ref{Fig : splitting with muxx muyy muxy}, we plot $\omega_{\mathrm{FM} } \sqrt{\Delta \mu_{xx} }$, $\omega_{\mathrm{FM} } \sqrt{ \Delta \mu_{yy} }$, and $\omega_{\mathrm{FM} } \sqrt{ \Delta \mu^{\prime}_{xy} }$ as functions of the external DC magnetic field. For the $xy$ component, $\omega$ in Eq.~(\ref{EQ : chi_xy}) can be approximated to $\omega_{\mathrm{FM} }$ because the splitting occurs around $\omega_{\mathrm{FM} }$, and we define $\Delta \mu^{\prime}_{xy}$ as
\begin{align} 
\Delta \mu^{\prime}_{xy} &= \Delta \mu_{xy} \frac{\gamma M }{\omega_{\mathrm{FM} } }. \label{EQ : muxy^prime}
\end{align}
These values monotonically increase with the increase of the magnetic field (see Fig.~\ref{Fig : splitting with muxx muyy muxy}). When damping $\alpha=0$, the frequency splitting is the light-matter coupling strength. In this polaritonic system, the coupling strength can be controlled by the magnetic field. Therefore, the results in Figure~\ref{Fig:transmission} can be explained as an interplay between the changing coupling strength and the matter loss (damping). As the magnetic field increases, the system evolves from weak coupling to strong coupling via an exceptional point. 

\bigskip
\section{Conclusion}
\label{Sec:Conclusion}
Overall, we have developed a microscopic theory for bulk magnon polaritons in YFeO$_3$ under magnetic fields. The model can be readily applied to any canted antiferromagnetic material supporting qFM and qAFM magnon modes.

\bibliography{si}